\documentclass[pre,10pt,amsmath,amssymb]{revtex4}
\usepackage{graphicx}

\usepackage{amsmath,amsfonts,amssymb,bm}

\begin{document}

\draft
\title{Anomalous diffusion in stochastic systems with nonhomogeneously distributed traps}

\author
{Tomasz Srokowski}

\affiliation{
 Institute of Nuclear Physics, Polish Academy of Sciences, PL -- 31-342
Krak\'ow,
Poland }

\date{\today}

\begin{abstract}
The stochastic motion in a nonhomogeneous medium with traps is studied and diffusion properties 
of that system are discussed. The particle is subjected to a stochastic stimulation 
obeying a general L\'evy stable statistics and experiences long rests due to 
nonhomogeneously distributed traps. The memory is taken into account by subordination of that process 
to a random time; then the subordination equation is position-dependent. The problem is 
approximated by a decoupling of the medium structure and memory and exactly solved 
for a power-law position dependence of the memory. In the case 
of the Gaussian statistics, the density distribution and moments are derived: 
depending on geometry and memory parameters, the system may reveal both the subdiffusion 
and enhanced diffusion. The similar analysis is performed for the L\'evy flights where 
the finiteness of the variance follows from a variable noise intensity near a boundary. 
Two diffusion regimes are found: in the bulk and near the surface. The anomalous diffusion 
exponent as a function of the system parameters is derived. 
\end{abstract} 

\pacs{05.40.Fb,02.50.-r}

\maketitle


\section{Introduction}

The diffusion is anomalous when the variance rises with time slower or faster than linearly and 
the density distribution differs from the normal distribution. 
This means that the central limit theorem is violated. One can expect that the theorem is not valid 
for transport processes if memory effects are present, e.g. when a medium contains traps. 
Traps hamper the transport and, as a consequence, 
a subdiffusion emerges -- as well as a stretched-Gaussian asymptotics of the distribution \cite{met}. 
We encounter such a situation for disordered media with impurities and defects. If the waiting time distribution 
is not prescribed to a given position and changes at each visit of the particle, we are dealing with 
an annealed disorder which corresponds to a renewal process. Then the continuous-time random walk (CTRW) 
is well suited to model the anomalous diffusion. 
If, on the other hand, the particle exercises a space structure which slowly evolves with time,  
the trapping time at a given site is the same for each visit of this site and a correlation between 
trapping times emerges (a quenched disorder) \cite{bou}. The quenched trap model \cite{ber,bar} 
takes into account that the particle may remember the rest time at a given 
point. Then the time-dependence of the variance 
can be derived by averaging over disorder \cite{bar} or using a renormalisation group approach \cite{mach}; 
it takes a form $\sim t^{2\beta/(1+\beta)}$, where $0<\beta<1$ characterises the memory. 

The central limit theorem may not apply also in the Markovian case since a nonhomogeneous structure 
of the environment makes subsequent random stimulations mutually dependent. 
Considering, in particular, the diffusion on fractals one must take into account 
the self-similar medium structure; 
it is described by the Fokker-Planck equation with a variable diffusion coefficient \cite{osh} (for a non-Markovian 
generalisation see \cite{met1,met2}). CTRW involves, in general, a coupled jump density distribution.  
When the waiting-time distribution is position-dependent, 
the Fokker-Planck equation, corresponding to the master equation, contains the variable diffusion 
coefficient \cite{kam} and, in the non-Markovian case, a variable order of the fractional derivative \cite{chevo}. 
For such inhomogeneous problems, CTRW implies a stretched-Gaussian shape of the density distribution 
and predicts the anomalous diffusion. 

If, on the other hand, the jump length does not obey the normal distribution 
but is governed by a general L\'evy stable distribution (L\'evy flights) the variance 
does not exist; such long jumps are frequently observed in many areas 
of science \cite{klag}. They may be directly related to a specific topology of the medium and then 
the transport description requires a variable diffusion coefficient: the folded polymers are a well-known 
example \cite{bro}. If a composite medium consists of many layers, the fractional equation 
is complicated and contains position-dependent both the diffusion coefficient and the order 
of the fractional derivative \cite{sti}. 
However, presence of the L\'evy flights does not need to imply infinite fluctuations because 
any physical system is finite and, if one introduces a truncation of the distribution tail, 
the diffusion properties are well-determined. 
It has been recently demonstrated that cracking of heterogeneous materials 
reveals a slowly falling power-law tail of the local velocity distribution of the crack front \cite{tall} but, 
despite that, the authors were able to determine the variance due to the finiteness of the system. 

Systems with memory can be conveniently handled in terms of a Langevin equation by introducing 
an auxiliary operational time. Process given by this equation is subsequently subordinated to 
the random physical time by means of a one-sided probability distribution with a long tail \cite{pir,zas}; 
such a system of the Langevin equations is used as an alternative formulation of CTRW \cite{fog,bar1} 
and also applied to quenched random media \cite{den} where the intensity of the random time distribution 
depends on the position. CTRW was applied to transport in the heterogeneous media by using 
fractional derivatives \cite{berk} and a subordinated system of Langevin equations \cite{berk1}. 
A subordination formalism, which directly takes into account that the memory in nonhomogeneous 
systems must depend on the position, has been recently proposed \cite{sro14}. This dependence was introduced 
via a variable intensity of the random time distribution and models the influence 
of the trap geometry on the local time lag. The system is then described by a set of two Langevin equations, 
\begin{eqnarray}
\label{la}
dx(\tau)&=&\eta(d\tau)\nonumber\\
dt(\tau)&=&g(x)\xi(d\tau), 
\end{eqnarray} 
where a non-negative random time intensity $g(x)$ defines a position-dependence of the memory effects and 
results from the presence of traps. 
Increments of the white noise $\eta$ are determined, in general, by the $\alpha$-stable 
L\'evy distribution, $L_\alpha(x)$, and $\xi(d\tau)$ stands for a stochastic process given by 
a one-sided distribution $L_\beta(x)$. Let us consider for the moment 
a Markovian case for which the density of $\xi$, 
still being one-sided, has finite moments (e.g. an exponential). Approximating $\xi$ by its mean, 
$\xi\to \langle\xi\rangle$, allows us to evaluate the operational time increment, 
$\Delta\tau=(\langle\xi\rangle g(x))^{-1}\Delta t$. The first equation (\ref{la}) can be discretized 
as $\Delta x=\Delta\eta\Delta\tau^{1/\alpha}$ and the system (\ref{la}) resolves itself 
to a single equation with a multiplicative noise,  
$dx(t)=\nu(x)^{1/\alpha}\eta(dt)$, where $\nu(x)=(\langle\xi\rangle g(x))^{-1}$. 
Since the $x-$dependence is evaluated from the other equation, $x(t)$ corresponds to the initial time 
for each interval and the It\^o interpretation is appropriate. 
The above equation incorporates the medium structure but neglects the memory effects which are important 
if $\beta<1$. Those effects, in turn, can be approximately taken into account by the subordination of the process 
$x(\tau)$ to a random time $t$ and, after that decoupling of the medium structure and memory, 
Eq.(\ref{la}) takes the form 
\begin{eqnarray}
\label{las}
dx(\tau)&=&\nu(x)^{1/\alpha}\eta(d\tau)\nonumber\\
dt(\tau)&=&\xi(d\tau);  
\end{eqnarray} 
Eq.(\ref{las}) refers to the case $\beta<1$ but $\nu(x)$ contains $\langle\xi\rangle$ corresponding 
to a Markovian process. 
The first part of Eq.(\ref{las}) describes such a process: a Markovian CTRW where the jump length is governed by 
$\eta(\tau)$ and the waiting time is Poissonian with a position-dependent rate $\nu(x)$. The derivation 
of this equation is presented in Appendix A. The system similar to (\ref{las}) (but including 
a potential) was considered in \cite{bar1}. It was demonstrated in Ref.\cite{sro14} that 
Eq.(\ref{las}) corresponds to a fractional kinetic equation \cite{zas} containing two fractional operators 
and a position-dependent diffusion coefficient. 
The anomalous transport predicted by Eq.(\ref{las}) is described by the variance for the Gaussian case and 
by the fractional moments for $\alpha<2$ \cite{sro14}. 

In the present paper, we analyse consequences on the diffusion process which follow from the dependence 
of memory on the position, defined by Eq.(\ref{la}). In the case of the L\'evy flights, we construct a process 
characterised by a finite variance taking into account that the system is finite. 
The finite variance emerges as a result of a variable noise intensity near a boundary and 
which decreases with the distance. This procedure differs from the usual noise truncation because 
the distribution of $\eta$ remains unaffected. 
The paper is organised as follows. In Sec.II we discuss the case $\alpha=2$ exactly solving Eq.(\ref{las}) 
for the power-law $g(x)$. 
Sec.III is devoted to the L\'evy flights. We discuss a system with the multiplicative noise 
in various interpretations and infer its general properties (Sec.IIIA). In Sec.IIIB, the multiplicative noise 
serves to model a boundary layer; we derive the variance, which then exists, 
as a function of time and all system parameters.

\section{Gaussian case} 

We consider a stochastic motion inside a medium containing traps and the particle 
is subjected to a white noise the increments of which obey 
the Gaussian statistics. The substrate nonhomogeneity is restricted to the time-characteristics 
of the system, determined by $g(x)$, and the dynamics is described by Eq.(\ref{la}). 
We assume the function $g(x)$ in a power-law form, 
\begin{equation}
\label{nuodx}
g(x)\sim|x|^\theta ~~~~~~(\theta>-1). 
\end{equation}
This form of the diffusion coefficient was assumed to describe, beside a problem of the diffusion 
on fractals, e.g. a turbulent two-particle diffusion \cite{fuj} and transport of fast electrons 
in a hot plasma \cite{ved}. It is encountered in geology where a power-law distribution of fracture 
lengths is responsible for transport in a rock \cite{pai}; this self-similar structure of 
a fracture and fault network is characterised by a fractal dimension and determines a pattern of water 
and steam penetration in the rock \cite{taf}. 
In Eq.(\ref{nuodx}), a negative $\theta$ means that intensity of the random time distribution 
is largest near the origin 
and diminishes with the distance; it rises with the distance for a positive $\theta$. 
We look for the density distribution of the particle position, $p(x,t)$, by solving Eq.(\ref{las}). 
The first equation (\ref{las}) in the It\^o interpretation leads to the Fokker-Planck equation 
\begin{equation}
\label{frace}
\frac{\partial p_0(x,\tau)}{\partial \tau}=
\frac{\partial^2[\nu(x) p_0(x,\tau)]}{\partial x^2},
\end{equation} 
which has the solution \cite{hen}
\begin{equation}
\label{sol20}
p_0(x,\tau)={\cal N}\tau^{-\frac{1+\theta}{2+\theta}} |x|^\theta\exp(-|x|^{2+\theta}/(2+\theta)^2\tau) 
\end{equation} 
(${\cal N}$ denotes a normalisation constant), and corresponds to the Markovian process $x(\tau)$. 
The random time, given by the second equation 
(\ref{las}), is defined by a one-sided, maximally asymmetric stable L\'evy distribution 
$L_\beta(\tau)$, where $0<\beta<1$, and the inverse distribution we denote by $h(\tau,t)$. 
The density $p(x,t)$ results from the integration of the densities $p_0(x,\tau)$ and $h(\tau,t)$ 
over the operational time, 
\begin{equation}
\label{inte}
p(x,t)=\int_0^\infty p_0(x,\tau)h(\tau,t)d\tau. 
\end{equation}
To evaluate the integral, it is convenient to use the Laplace transform from Eq.(\ref{inte}) 
taking into account that $\bar h(\tau,u)=u^{\beta-1}\exp(-\tau u^\beta)$ \cite{wer}. 
Then a direct integration yields a Laplace transform from the normalised density, 
\begin{equation}
\label{lbes}
\bar p(x,u)=-2\frac{(2+\theta)^\nu}{\Gamma(-\nu)}|x|^{\theta+1/2}u^c
{\mbox K}_\nu\left(\frac{|x|^{1+\theta/2}}{1+\theta/2}u^{\beta/2}\right), 
\end{equation}  
where K$_\nu(z)$ is a modified Bessel function, $\nu=1/(2+\theta)$ and 
$c=\beta-\beta/(4+2\theta)-1$. 
The density obtained from the inversion of the above transform can be expressed 
in terms of the Fox H-function,  
\begin{eqnarray} 
\label{sol2}
p(x,t)=-\frac{2}{\beta t}\frac{(1+\theta/2)^{\nu+2c/\beta}}{\Gamma(-\nu)}|x|^{(2+\theta)/\beta-1}
H_{1,2}^{2,0}\left[\frac{|x|^{(2+\theta)/\beta}}{(2+\theta)^{2/\beta}t}\left|\begin{array}{l}
~~~~~~~~~~~~~~~~~~~(0,1)\\
\\
(c/\beta-\nu/2,1/\beta),(c/\beta+\nu/2,1/\beta)
\end{array}\right.\right]; 
\end{eqnarray} 
details of the derivation are presented in Appendix B. 
Eq.(\ref{sol2}) seems complicated but an expression for large $|x|$ is simple: it has 
a stretched-Gaussian form which follows from an asymptotic formula for the H-function \cite{bra}, 
\begin{equation}
\label{strg2}
p(x,t)\sim |x|^{\theta/\beta}t^{-\frac{\beta(1+\theta)}{(2+\theta)(2-\beta)}}
\exp\left[-A|x|^{(2+\theta)/(2-\beta)}/t^{\frac{\beta}{2-\beta}}\right], 
\end{equation}
where $A=(2/\beta-1)/[\beta^{3/(2-\beta)}(2+\theta)^{2/(2-\beta)}]$. The stretched-Gaussian shape 
of the tail is typical for diffusion on fractals \cite{met2} and emerges in the trap models \cite{ber}. 

We are interested in the moments of $p(x,t)$; all of them are finite and given by 
a characteristic function which can be derived as a series expansion. Eq.(\ref{inte}) yields 
\begin{equation}
\label{chf1}
\widetilde p(k,t)=\sum_{n=0}^\infty\frac{(-1)^n}{(2n)!}k^{2n}\int_0^\infty\langle x^{2n}\rangle_{p_0}h(\tau,t)d\tau,
\end{equation}
where the moments of $p_0(x,\tau)$ directly follow from Eq.(\ref{sol20}), 
\begin{equation}
\label{chf2}
\langle x^n\rangle_{p_0}=-(2+\theta)^{2n/(2+\theta)+1}\frac{\Gamma[(1+n+\theta)/(2+\theta)]}{\Gamma[-1/(2+\theta)]}
\tau^{n/(2+\theta)}. 
\end{equation}
Then the integral resolves itself to the moments of $h(\tau,t)$ that are given by \cite{pir} 
\begin{equation}
\label{chf3}
\langle\tau^{2n/(2+\theta)}\rangle_h=\frac{\Gamma[2n/(2+\theta)+1]}{\Gamma[2n\beta/(2+\theta)+1]} 
t^{2n\beta/(2+\theta)}
\end{equation}
and the final expression for the characteristic function reads 
\begin{equation}
\label{chf}
\widetilde p(k,t)=-\frac{2+\theta}{\Gamma[-1/(2+\theta)]}
\sum_{n=0}^\infty\frac{(-1)^n}{(2n)!}(2+\theta)^{4n/(2+\theta)}\Gamma[(1+2n+\theta)/(2+\theta)]
\frac{\Gamma[2n/(2+\theta)+1]}{\Gamma[2n\beta/(2+\theta)+1]} t^{2n\beta/(2+\theta)}k^{2n}. 
\end{equation}
Diffusion properties are determined by the variance the time-dependence of which follows from 
scaling arguments or from Eq.(\ref{chf}): 
$\langle x^2\rangle(t)=-\partial^2\widetilde p(k=0,t)/\partial k^2 \sim t^{2\beta/(2+\theta)}$; 
this formula indicates both the sub- and superdiffusion. 
Comparison of the above approximate result with a numerical solution of Eq.(\ref{la}) reveals 
a reasonable agreement; some discrepancies emerge for a small $\beta$ and large $\theta$ \cite{sro14}. 

Finally, we will demonstrate that $p(x,t)$, Eq.(\ref{sol2}), satisfies the fractional equation 
\begin{equation}
\label{glef}
\frac{\partial p(x,t)}{\partial t}={_0}D_t^{1-\beta}\frac{\partial^2}{\partial x^2}(|x|^{-\theta} p(x,t)), 
\end{equation} 
where ${_0}D_t^{1-\beta}$ is a Riemann-Liouville operator 
\begin{equation}
\label{rlo}
_0D_t^{1-\beta}f(t)=\frac{1}{\Gamma(\beta)}\frac{d}{dt}\int_0^t dt'\frac{f(t')}{(t-t')^{1-\beta}},  
\end{equation} 
and which equation constitutes a non-Markovian generalisation of a Fokker-Planck equation for CTRW 
with a position-dependent waiting-time distribution \cite{sro06}. First, we integrate Eq.(\ref{glef}) 
over time, 
\begin{equation}
\label{glef1}
p(x,t)-p_0(x)={_0}D_t^{-\beta}\frac{\partial^2}{\partial x^2}(|x|^{-\theta} p(x,t)), 
\end{equation} 
where $p_0(x)$ is the initial condition, and take the Laplace transform, 
\begin{equation}
\label{glef2}
u^\beta\bar p(x,u)-u^{\beta-1}p_0(x)=\frac{\partial^2}{\partial x^2}(|x|^{-\theta} \bar p(x,u)). 
\end{equation} 
The differentiation produces a differential equation 
\begin{equation}
\label{glef3}
x^2\bar p''-\theta x\bar p'+[\theta(1+\theta)-u^\beta x^{\theta+2}]\bar p+u^{\beta-1}x^{\theta+2}p_0=0 
\end{equation}
and the last term vanishes if $p_0(x)=\delta(x)$. Its particular solution has the form \cite{kamke} 
\begin{equation}
\label{glef4}
\bar p(x,u)=|x|^{\theta+1/2}
{\mbox K}_\nu\left(\frac{|x|^{1+\theta/2}}{1+\theta/2}u^{\beta/2}\right)f(u), 
\end{equation}  
where $f(u)$ is an arbitrary function. Putting $f(u)\sim u^c$, we obtain Eq.(\ref{lbes}). 

The power-law form of $g(x)$ suggests an interpretation of the memory position-dependence 
as a fractal trap structure. In this picture, $g(x)$ means a density of traps, 
i.e. the number of traps per unit interval, which is self-similar. 
The density of a fractal embedded in the one-dimensional space equals $|x|^{d_f-1}$, where 
$d_f$ stands for the fractal (Hausdorff) dimension \cite{hav}. Comparing the above expression with $g(x)$ 
given by Eq.(\ref{nuodx}) allows us to interpret the parameter $\theta$ by relating it to the 
fractal dimension: $\theta=d_f-1$ for $\theta\in(-1,0]$. Then the expression for the variance takes the form, 
\begin{equation}
\label{warg}
\langle x^2\rangle(t)\sim t^{2\beta/(1+d_f)}. 
\end{equation} 
The subdiffusion, observed in the homogeneous case, may turn to the enhanced diffusion if 
dimension of the trap structure is small and the memory sufficiently weak (large $\beta$). 
The case $\beta=1$ is special: $\langle\xi\rangle$ does not exist and, 
for $d_f=1$ ($\theta=0$), the problem resolves itself 
to the ordinary one-dimensional CTRW which is characterised by a weak subdiffusion, 
$\langle x^2\rangle(t)\sim t/\ln t$ \cite{bou}. 

The present problem differs from the random walk on fractals when the particle performs jumps 
according to a self-similar pattern \cite{osh}. 
The non-Markovian generalisation of that model \cite{met1} contains a fractional equation which has a form 
different from Eq.(\ref{glef}) and the predicted motion is always subdiffusive. 
The density has a finite value at the origin in contrast to the our approach: $p(x,t)$, 
Eq.(\ref{sol2}), diverges at $x=0$ ($\theta<0$) where the trapping is strong and the evolution, as a function 
of the physical time, proceeds very slowly near the origin. 

\section{L\'evy flights}

Motion of a massive particle subjected to an additive noise which obeys the L\'evy statistics with long jumps 
is characterised by the infinite variance; this situation is unacceptable for physical reasons. However, 
the additive noise is an idealisation and in realistic systems the stochastic 
stimulation may depend on a state of the system. For this -- multiplicative -- noise, the variance 
may be finite and the anomalous diffusion exponent well determined. 
In the next subsection, we discuss general properties of 
the non-Markovian Langevin dynamics with that noise; the memory effects are taken into account 
by a subordination of the dynamical process to the random time. One can expect 
that the multiplicative noise emerges near a boundary where the environment structure is more 
complicated than in the bulk. The boundary effects modelled in terms of the multiplicative 
noise will be discussed in Subsection IIIB. 

\subsection{Multiplicative noise}

The generalisation of Eq.(\ref{la}) including the multiplicative noise is the following 
\begin{eqnarray}
\label{lam}
dx(\tau)&=&f(x)\eta(d\tau)\nonumber\\
dt(\tau)&=&g(x)\xi(d\tau) 
\end{eqnarray} 
and, in the case of the first equation (\ref{lam}), we must decide how this equation is to be interpreted, 
namely at which time $f(x(t))$ is to be evaluated. More precisely, one defines 
a stochastic integral as a Riemann integral, 
\begin{equation}
\label{riem}
\int_0^t f[x(\tau)]d\eta(\tau)=
\sum_{i=1}^n f[(1-\lambda_I)x(\tau_{i-1})+\lambda_Ix(\tau_i)][\eta(\tau_i)-\eta(\tau_{i-1})], 
\end{equation} 
where the interval $(0,t)$ has been divided in $n$ subintervals ($n\to\infty$). 
The parameter $0\le\lambda_I\le1$ determines the interpretation and corresponds, in particular, 
to the It\^o (II) ($\lambda_I=0$), Stratonovich (SI) ($\lambda_I=1/2$) 
and anti-It\^o interpretation (AII) ($\lambda_I=1$). II applies, in particular, if the noise consists 
of clearly separated pulses, e.g. for a continuous description of integer processes and is used 
in the perturbation theory due to its simplicity. It is well-known that for the 
Gaussian processes the ordinary rules of the calculus are valid in the case of SI, in contrast to 
the other interpretations, which allows us to transform the equation with the multiplicative noise into 
an equation with the additive noise by a simple variable change. As regards the general stable processes, 
we observe the same property: 
the numerical solutions of the Langevin equation by using Eq.(\ref{riem}) with $\lambda_I=1/2$ agree 
with those where a standard technique of the variable change is applied \cite{sro09}. 
Obviously, the ordinary rules of the calculus are always valid if one defines 
the white noise as a limit of the coloured noise \cite{sro12}; equivalence of that limit and SI 
is an important property of the multiplicative Gaussian processes. 
In the case of Eq.(\ref{lam}), we obtain the equation 
with the additive noise by changing the variable, $y(x)=\int_0^x\frac{dx'}{f(x')}$, in the first 
of Eq.(\ref{lam}). Next, we decouple the medium structure and memory (cf. Eq.(\ref{las})) and get the equation 
$dy(\tau)=\nu(x(y))^{1/\alpha}\eta(d\tau)$. It contains a multiplicative noise in II 
and corresponds to the Fokker-Planck equation \cite{sche} 
\begin{equation}
\label{fp0}
\frac{\partial p_0(y,\tau)}{\partial \tau}=
\frac{\partial^\alpha[\nu(x(y)) p_0(y,\tau)]}{\partial|y|^\alpha}.
\end{equation} 
Solution in the original variable $x$ directly follows from the solution of Eq.(\ref{fp0}), 
\begin{equation}
\label{solx}
p_0(x,\tau)=\frac{1}{f(x)}p_0(y(x),\tau),  
\end{equation}
and the density as a function of the physical time -- determined by the equation $dt(\tau)=\xi(d\tau)$ -- 
results from Eq.(\ref{inte}). 

In the following, we restrict our considerations to a power-law form of $g(x)$: 
\begin{equation}
\label{nul}
\nu(x)=|x|^{-\theta\alpha} ~~~~~~(\theta>-1), 
\end{equation}
similar to Eq.(\ref{nuodx}). Then the waiting time is either large near the origin corresponding 
to a large intensity of the random time density ($\theta<0$) or the probability of long rests rises with the distance 
($\theta>0$). Eq.(\ref{fp0}) can be solved if one neglects terms higher than $|k|^\alpha$ 
in the characteristic function \cite{sro06,sro14a} which, after inverting the Fourier transform, 
yields tail of the density (\ref{solx}): $p_0(y,\tau)\sim |y|^{-1-\alpha}$; it corresponds 
to the L\'evy-stable asymptotics \cite{uwa}. The backward transformation of the variable, 
$y\to x$, produces the asymptotics $p_0(x,\tau)\sim f(x)^{-1}y(x)^{-1-\alpha}$, 
which indicates that finite moments can exist. In particular, the variance is finite 
if $f(x)$ satisfies the condition 
\begin{equation}
\label{warvar}
\lim_{x\to\infty}\frac{x^3}{f(x)y(x)^{1+\alpha}}=0. 
\end{equation}
We assume from now on that $f(x)$ has the power-law form, 
\begin{equation}
\label{fodx}
f(x)=|x|^{-\gamma}, 
\end{equation}
which allows us to obtain exact solutions. Then Eq.(\ref{warvar}) 
implies the condition for the finite variance: $\gamma>2/\alpha-1$. 
Inserting the functions $g(x)$ and $f(x)$, given by Eq.(\ref{nul}) and Eq.(\ref{fodx}), 
to Eq.(\ref{lam}) we obtain after elimination of the position-dependent factor in the subordination 
equation and straightforward calculations the following set of the Langevin equations, 
\begin{eqnarray}
\label{lam1}
dy(\tau)&=&D|y|^{-\theta/(1+\gamma)}\eta(d\tau)\nonumber\\
dt(\tau)&=&\xi(d\tau), 
\end{eqnarray} 
where $D=(1+\gamma)^{-\theta/(1+\gamma)}$. The first of the above equations corresponds to the Fokker-Planck equation, 
\begin{equation}
\label{fp01}
\frac{\partial p_0(y,\tau)}{\partial \tau}=
D^\alpha\frac{\partial^\alpha[|y|^{-\alpha\theta/(1+\gamma)} 
p_0(y,\tau)]}{\partial|y|^\alpha}
\end{equation} 
which in the limit of small wave numbers is satisfied by a density determined by the following 
characteristic function in respect to the variable 
$y=\frac{1}{1+\gamma}|x|^{1+\gamma}\mbox{sign }x$, 
\begin{equation}
\label{py0k}
\widetilde p_0(k,\tau)\approx 1-(A_0\tau)^{c_\theta}|k|^\alpha; 
\end{equation}
in the above equation, $c_\theta=1/(\alpha+\theta/(1+\gamma))$ and 
\begin{equation}
\label{staa}
A_0=\frac{2D(1+\theta+\gamma)}{\pi\alpha(1+\gamma)}
\Gamma(\theta/(1+\gamma))\Gamma(1-\alpha\theta/(1+\gamma))
\sin\left(\frac{\pi\alpha\theta}{2(1+\gamma)}\right). 
\end{equation}
Eq.(\ref{py0k}) corresponds to the stable density $L_\alpha$ if 
$\theta\in(-(1+\gamma),(1+\gamma)/\alpha)$ \cite{sro14a}. 
The final density $p(x,t)$ results from the transformation to the physical time by means 
of Eq.(\ref{inte}) which we rewrite as a Fourier transform in respect to $y$. Using Eq.(\ref{py0k}) yields 
\begin{equation}
\label{intek}
\widetilde p(k,t)=1-A_0^{c_\theta}|k|^\alpha\int_0^\infty \tau^{c_\theta} h(\tau,t)d\tau 
\end{equation}  
and the integral can be easily evaluated if we express the function $h(\tau,t)$ by the H-function \cite{non}, 
\begin{eqnarray} 
\label{htt}
h(\tau,t)=\frac{1}{\beta\tau}H_{1,1}^{1,0}\left[\frac{\tau^{1/\beta}}{t}\left|\begin{array}{l}
(1,1)\\
\\
(1,1/\beta)
\end{array}\right.\right].  
\end{eqnarray} 
Changing the variable in the integral (\ref{intek}), $\xi=\tau^{1/\beta}/t$, we get 
\begin{equation}
\label{intek1}
\widetilde p(k,t)=1-A_0^{c_\theta}|k|^\alpha t^{\beta c_\theta}\int_0^\infty \xi^{\beta c_\theta-1}H(\xi)d\xi, 
\end{equation}  
where $H(\xi)$ means the H-function in Eq.(\ref{htt}). We evaluate the resulting Mellin transform 
and obtain the Fourier transform corresponding to the stable distribution for small $|k|$. 
Inversion of this Fourier transform produces the final expression, 
\begin{equation}
\label{px}
p(x,t)=A^{-1} t^{-\beta c_\theta}|x|^\gamma L_\alpha(t^{-\beta c_\theta}|y(x)|/A),
\end{equation}
where we applied Eq.(\ref{solx}) and denoted 
\begin{equation}
\label{staa1}
A=A_0^{c_\theta/\alpha}\Gamma[1+c_\theta]^{1/\alpha}/\Gamma[1+\beta c_\theta]^{1/\alpha}.
\end{equation} 
The solution (\ref{px}) has the asymptotics 
\begin{equation}
\label{asyt}
p(x,t)\sim t^{\beta c_\theta}|x|^{-1-\alpha-\alpha\gamma}. 
\end{equation}
\begin{center}
\begin{figure}
\includegraphics[width=95mm]{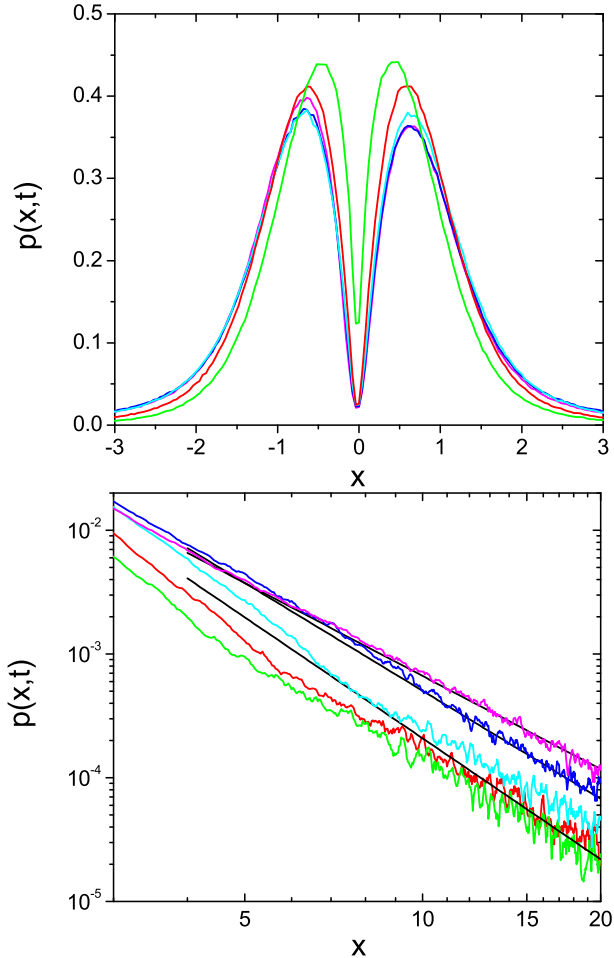}
\caption{(Colour online) Distributions obtained from a numerical integration of Eq.(\ref{lam}) with 
$g(x)=|x|^{\theta\alpha}$ and $f(x)=|x|^{-\gamma}$ for $\alpha=3/2$, $\theta=2/3$, $\beta=1/2$, $\gamma=1/2$ and $t=0.1$. 
Curves correspond to the following 
values of $\lambda_I$: 0 (magenta), 0.1 (blue), 0.2 (cyan), 0.5 (red) and 1 (green) (from bottom 
to top in the upper part and from top to bottom in the lower part). The straight lines correspond to 
a power-law function with the index 2.5, 2.9 and 3.25 (from top to bottom).}
\end{figure}
\end{center} 
The variance can be directly evaluated from (\ref{px}): 
\begin{equation}
\label{warm}
\langle x^2\rangle(t)=2\int_0^\infty x^2 p(x,t)dx=2A^{-1}t^{-\beta c_\theta}
\int_0^\infty x^{2+\gamma}L_\alpha(A^{-1}(1+\gamma)^{-1}t^{-\beta c_\theta}x^{1+\gamma})dx. 
\end{equation}
Change of the variable and evaluating the Mellin transform yields the final expression, 
\begin{equation}
\label{warmf}
\langle x^2\rangle(t)=-\frac{2}{\pi\alpha}[A(1+\gamma)]^{2/(1+\gamma)}\Gamma\left(-\frac{2}{\alpha(1+\gamma)}\right)
\Gamma\left(1+\frac{2}{1+\gamma}\right)\sin\left(\frac{\pi}{1+\gamma}\right)t^{2\beta c_\theta/(1+\gamma)}, 
\end{equation}
which means that motion is always subdiffusive. 

On the other hand, if one understands the noise $\eta$ in Eq.(\ref{lam}) in a sense of II, the Langevin 
equation for $x(\tau)$ contains a multiplicative term $|x|^{-\theta-\gamma}$ and 
the counterpart of Eq.(\ref{fp01}) reads 
\begin{equation}
\label{fp01i}
\frac{\partial p_0(x,\tau)}{\partial \tau}=
\frac{\partial^\alpha[|x|^{-\alpha(\theta+\gamma)} 
p_0(x,\tau)]}{\partial|x|^\alpha}. 
\end{equation} 
Solving the above equation in the diffusion limit and evaluation of the integral (\ref{inte}) 
produces the distribution $p(x,t)$ with the asymptotics $\sim|x|^{-1-\alpha}$. 
Comparison with Eq.(\ref{asyt}) indicates a qualitative difference between II and SI for the L\'evy flights: 
whereas for II presence of the multiplicative factor in the noise influences only the time-dependence, 
for SI it modifies the tail shape and makes finiteness of the variance possible. 
In the face of that difference, it is interesting to check what 
the density distributions for the other interpretations look like. To find those distributions, 
we have to resort to the numerical analysis. According to Eq.(\ref{riem}), the discretized form 
of the Langevin equation for the arbitrary interpretation is given by the following expression, 
\begin{equation}
\label{riem1}
x_{n+1}=x_n+[(1-\lambda_I)x_n+\lambda_I x_{n+1}]^{-\gamma}\eta_n h^{1/\alpha},
\end{equation}
where $\eta_n$ is sampled from a symmetric stable distribution according to a well-known 
algorithm \cite{wer1} and $h$ is a time step. The expression for $x_{n+1}$ is not explicit and Eq.(\ref{riem1}) can be 
exactly solved only for a few values of $\gamma$; in general, it must solved numerically at every integration step. 
For that purpose, we applied the parabolic interpolation scheme (the Muller method) \cite{ral}. 
A simple modification of the standard method \cite{wer} allows us to evaluate 
the variable $x$ as a function of the physical time -- determined by the second equation (\ref{lam}) -- 
without an explicit derivation of the subordinator $t(\tau)$. 
Examples of distributions for a few values of $\lambda_I$ are presented in Fig.1, separately 
for the central part and for the tails. We observe a similar slope of the tail for all $\lambda_I>0.2$; in particular,  
it is almost identical for SI and AII. Near the origin, in turn, some differences emerge and the height of the peak 
rises with $\lambda_I$. For all the interpretations, $p(0,t)=0$ which is a consequence of the divergence 
of $f(x)$ in the origin.

\subsection{Boundary effects and anomalous diffusion} 

Let us assume that the particle is subjected to a noise the intervals of which are governed by the symmetric 
L\'evy stable distribution. The transport proceeds inside a medium with traps and intensity of the 
random time density 
is given by the function $g(x)$; the dynamics is described by the Langevin equation with the additive noise, 
Eq.(\ref{la}). Such systems are characterised by the infinite variance which, for massive particles, 
violates physical principles and attempts were undertaken to suppress long tails in the L\'evy density. 
A simple remedy is to introduce a modification of the stable distribution to make the tail 
steeper. Such a truncation may be assumed as a simple cut-off \cite{man} or involve some rapidly falling 
function: e.g. an exponential \cite{kop} or a power-law $|x|^{-\beta}$, 
where $\beta\ge2-\alpha$ \cite{sok}. Processes involving the truncated distributions actually 
converge to the normal distribution, according to the central limit theorem, but 
the power-law tails may be visible for a long time due to a slow convergence. 
On the other hand, variance becomes finite if one takes into account a finiteness of 
the particle velocity (L\'evy walk) \cite{met}. 
In the present approach, the finite variance results from a variable intensity of the noise in 
the region close to the boundary whereas intervals of the noise are always distributed according to 
the stable distribution. As a consequence, the Langevin equation acquires a multiplicative noise and 
the requirement of the finiteness of the variance imposes a condition on the form of the position-dependent 
noise intensity. We assume, in addition, that the boundary effects do not affect the trap structure. 
Presence of that noise is natural: one can expect that the additional complication of the environment structure 
in the vicinity of the boundary introduces a dependence of the noise on the process value and 
requires a more general approach than an equation with the additive noise. 
Emergence of the multiplicative noise near a boundary has been experimentally demonstrated for some physical systems. 
For example, a description of the colloidal particles diffusion in terms of a constant diffusion coefficient
appears possible only if a particle remains far from any boundary \cite{brett}. 
Moreover, presence of the multiplicative noise in a description of particles near a wall 
is necessary to reach a proper thermal equilibrium \cite{lau}. 

We assume that the boundary effects become important at a distance $|x|=L$ and then the Langevin 
equation acquires the multiplicative noise. Its intensity is parametrised by $f(x)$ as a falling 
power-law function, 
\begin{eqnarray} 
\label{fodxtr}
f(x)=\left\{\begin{array}{ll}
1  &\mbox{for  $|x|\le L$}\\
L^\gamma|x|^{-\gamma} &\mbox{for  $|x|>L$},
\end{array}
\right.
\end{eqnarray} 
where $\gamma>0$; the dynamics is governed by Eq.(\ref{lam}) and 
the noise in the first equation will be interpreted according to SI. 
A new variable, 
\begin{eqnarray} 
\label{trxy}
y(x)=\left\{\begin{array}{ll}
x  &\mbox{for  $|x|\le L$}\\
 \frac{L}{1+\gamma}\left[\gamma+(|x|/L)^{1+\gamma}\right)\mbox{]sign }x &\mbox{for  $|x|>L$},
\end{array}
\right.
\end{eqnarray} 
allows us to get rid of the multiplicative factor $f(x)$ in Eq.(\ref{lam}) 
and the equation corresponding to the first equation (\ref{lam1}) takes the form 
\begin{eqnarray}
\label{lamtr}
dx(\tau)&=&|x|^{-\theta}\eta(d\tau)\mbox{\hskip6cm for  $|x|\le L$}\nonumber\\
dy(\tau)&=&[(|y|-L)(1+\gamma)L^\gamma+L^{1+\gamma}]^{-\theta/(1+\gamma)}\eta(d\tau)\mbox{\hskip13mm for  $|x|>L$.} 
\end{eqnarray} 
In general, the density distribution can be obtained in a closed form only for $|x|\le L$ and $|x|\gg L$. 
We consider, at the beginning, the case $\theta=0$ corresponding to a uniform trap distribution and 
follow a similar method as in the preceded subsection: solve the 
Fokker-Planck equation and integrate over the operational time.  The final result for $|x|>L$ reads 
\begin{equation} 
\label{soltr}
p(x,t)=A^{-1}t^{-\beta/\alpha}L^{-\gamma}
|x|^\gamma L_\alpha(t^{-\beta/\alpha}|y(x)|/A),
\end{equation} 
where $A=\Gamma(1+\beta)^{-1/\alpha}$. 
The asymptotic form of the above equation, $p(x,t)\sim|x|^{-1-\alpha-\alpha\gamma}$, reveals 
the meaning of the parameter $\gamma$: the limit $\gamma\to\infty$, for which $f(x)=1-\Theta(|x|-L)$, 
yields $p(x,t)\to 0$ ($|x|>L$), 
i.e. $x=\pm L$ becomes an absorbing barrier and the surface is reduced to single points. For a finite value 
of $\gamma$, the system is not strictly confined but the probability density of finding 
the particle in the outer region rapidly decreases with the distance if $\gamma$ is large. 
Moreover, for a large $L$, a very long time is needed to observe the particle at distances markedly 
larger than $L$. 
The diffusion problem is well-defined since the finite variance exists if $\gamma>2/\alpha-1$ and, 
in the following, we will calculate the variance on this assumption. 
The system defined by Eq.(\ref{lamtr}) changes its properties at $|x|=L$ and one can expect 
a different diffusion behaviour in the surface region, compared to the bulk. 
If $L$ is sufficiently large, the dynamics resolves itself 
to the truncated L\'evy flights and the diffusion properties are equivalent to the Gaussian 
case \cite{trunc}, providing the time is relatively small; then one gets the standard anomalous 
diffusion law, 
\begin{equation}
\label{war1t0}
\langle x^2\rangle(t)\sim t^\beta. 
\end{equation}
Variance in the limit $t\to\infty$ follows from a direct evaluation of the integral. One can easily 
demonstrate that the contribution from $|x|<L$ is small for a large time. Then 
\begin{equation}
\label{war2t0}
\langle x^2\rangle(t)=\frac{2t^{-\beta/\alpha}}{AL^\gamma}\int_L^\infty x^{\gamma+2}
L_\alpha\left(\frac{|y(x)|}{At^{\beta/\alpha}}\right)dx
\end{equation}
and, introducing a new variable $x'=t^{-\beta/\alpha(1+\gamma)}x$, we obtain 
\begin{equation}
\label{ygr}
y=L+\frac{L^{-\gamma}}{1+\gamma}(x'^{1+\gamma}t^{\beta/\alpha}-L^{1+\gamma})\to 
\frac{L^{-\gamma}}{1+\gamma}x'^{1+\gamma}t^{\beta/\alpha}~~~~(t\to\infty)
\end{equation}
for any $x'$ which allows us to reduce Eq.(\ref{war2t0}) to the form 
\begin{equation}
\label{war2t01}
\langle x^2\rangle(t)=\frac{2t^{2\beta/\alpha(1+\gamma)}}{AL^\gamma}\int_{x_0}^\infty x^{\gamma+2}
L_\alpha\left(\frac{L^{-\gamma}}{A(1+\gamma)}x^{1+\gamma}\right)dx,
\end{equation}
where $x_0=L/t^{\beta/\alpha(1+\gamma)}\to 0$. The integral can be evaluated by applying the standard 
properties of the H-function and, after lengthy but straightforward calculations, we obtain the final 
formula, 
\begin{equation}
\label{war2t0f}
\langle x^2\rangle(t)=-\frac{2}{\pi\alpha}L^{\alpha\gamma c_\gamma}(1+\gamma)^{\alpha c_\gamma}
\Gamma(1+\beta)^{-c_\gamma}\Gamma(-c_\gamma)\Gamma(\alpha c_\gamma)\sin\left(\frac{\pi}{1+\gamma}\right)t^{\beta c_\gamma},
\end{equation}
where $c_\gamma=2/\alpha(1+\gamma)$. Since $c_\gamma<1$, the motion is always subdiffusive: the variance 
in the limit $t\to\infty$ rises with time slower than linearly and also slower than in the case 
of a small time, Eq.(\ref{war1t0}). The slope explicitly depends on $\alpha$, in contrast to Eq.(\ref{war1t0}), 
and drops to zero for a sharp edge ($\gamma\to\infty$). Therefore, 
we observe two diffusion regimes; the standard form of the variance (\ref{war1t0}), 
which is typical for the Gaussian case and the truncated L\'evy flights, 
emerges for relatively small times and corresponds to trajectories 
abiding not far from the origin. On the other hand, when at large time the surface region becomes 
important, diffusion is weaker. 
Those two diffusion regimes are illustrated in Fig.2 for the Cauchy distribution ($\alpha=1$) 
where the variance was obtained from $p(x,t)$ by a direct integral evaluation: slopes agree 
with Eq.(\ref{war1t0}) and Eq.(\ref{war2t0f}). 
\begin{center}
\begin{figure}
\includegraphics[width=95mm]{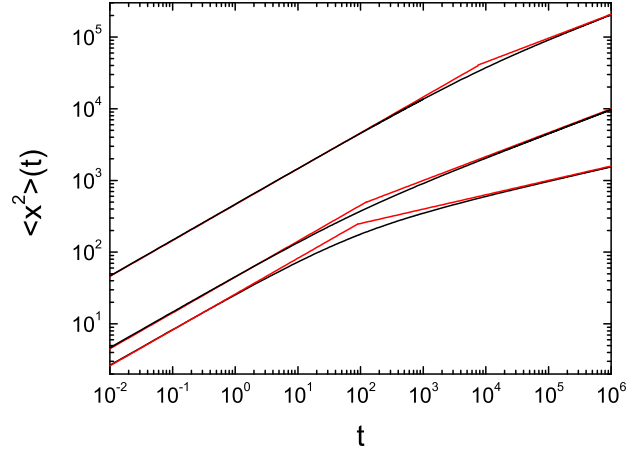}
\caption{(Colour online) Variance as a function of time for $\alpha=1$ and $\beta=1/2$. 
The curves (from bottom to top) correspond to: 1. $\gamma=4$ and $L=10$; 2. $\gamma=2$ and $L=10$; 
3. $\gamma=2$ and $L=100$. Straight-line segments (marked by the red lines) on the left-hand side 
have the slope 1/2 and those on the right-hand side: 1/5, 1/3 and 1/3.} 
\end{figure}
\end{center} 
\begin{center}
\begin{figure}
\includegraphics[width=95mm]{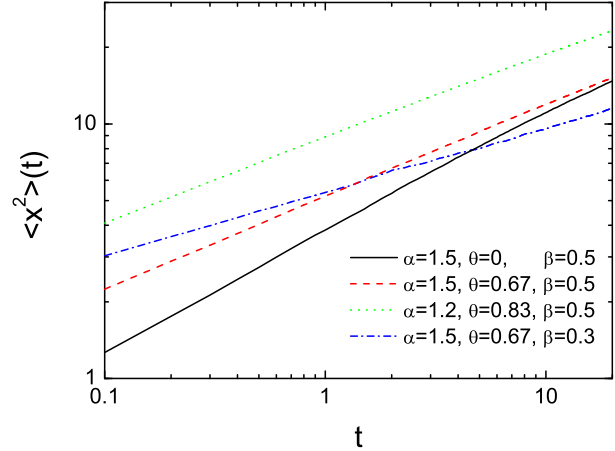}
\caption{(Colour online) Variance as a function of time for a few sets of the parameters 
$\alpha$, $\theta$ and $\beta$; the other parameters: $L=10$ and $\gamma=2$.} 
\end{figure}
\end{center} 
\begin{center}
\begin{figure}
\includegraphics[width=95mm]{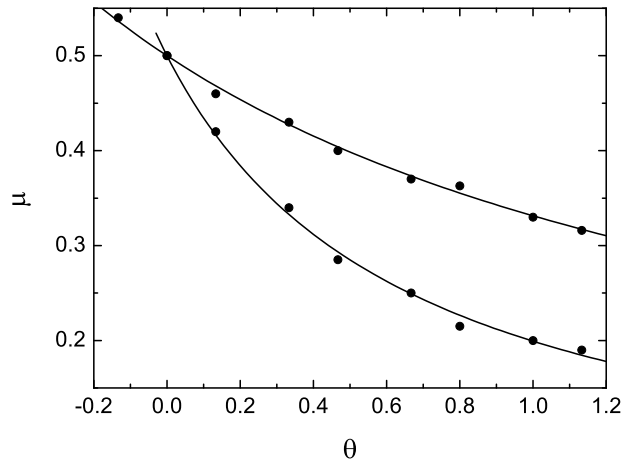}
\caption{Slope of the time-dependence of the variance, $t^\mu$, as a function of $\theta$ for $\beta=0.5$ 
and two values of $\alpha$: 0.5 (lower points) and 1.5 (upper points). The other parameters: $\gamma=2$ and $L=10$. 
Solid lines mark the function  $\mu=0.5/(1+c(\alpha)\theta)$, where $c(0.5)=1.51$ and $c(1.5)=0.51$.} 
\end{figure}
\end{center} 

\begin{center}
\begin{figure}
\includegraphics[width=95mm]{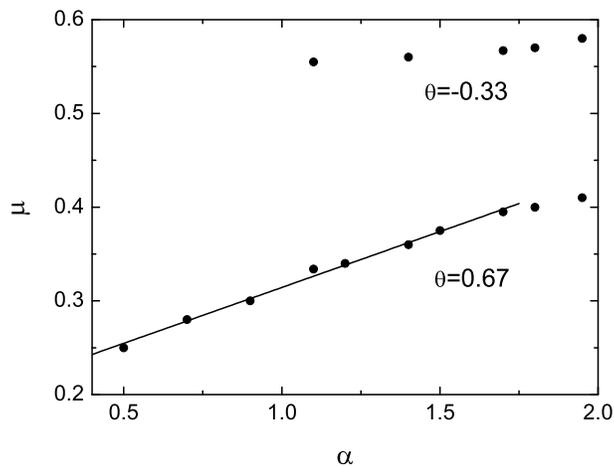}
\caption{Slope as a function of $\alpha$ for two values of $\theta$, the other parameters are the same as in Fig.4. 
Solid line marks the function $\mu=0.2+0.12\alpha$.} 
\end{figure}
\end{center} 
\begin{center}
\begin{figure}
\includegraphics[width=95mm]{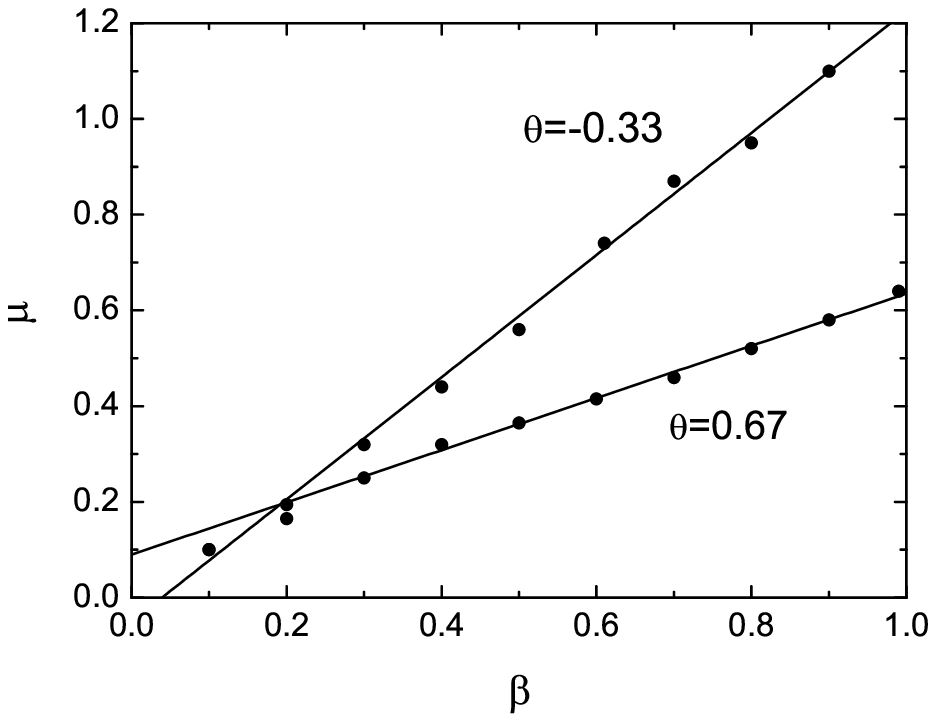}
\caption{Slope as a function of $\beta$ for two values of $\theta$, the other parameters are the same as in Fig.4. 
Solid lines mark the functions $\mu=0.09+0.545\beta$ and $\mu=-0.05+1.276\beta$ for the positive 
and negative $\theta$, respectively.} 
\end{figure}
\end{center} 

If $\theta\ne 0$, Eq.(\ref{lamtr}) is not manageable analytically except a non-physical case $|x|\gg L$. 
Then the variance was determined by a numerical solving of Eq.(\ref{lam}) where 
the multiplicative noise was treated according to Eq.(\ref{riem1}) with $\lambda_I=1/2$. 
The numerical calculations show that the time-dependence of the variance has the same 
form as Eq.(\ref{war1t0}) but with a modified index, $\sim t^\mu$. 
This form was found for all the parameters -- $\theta$, $\alpha$ and $\beta$ -- 
if $L$ was large. The above observation is illustrated in Fig.3 where some 
examples of $\langle x^2\rangle(t)$ are presented. 
Dependence of the slope on the parameters is presented in the subsequent figures. 
Fig.4 shows $\mu$ as a function of $\theta$ for two values of $\alpha$; it diminishes with $\theta$ 
and the numerical results reveal a dependence $\mu=0.5/(1+c(\alpha)\theta)$ (the coefficient 
$c(\alpha)$ is indicated in the figure). 
The dependence of $\mu$ on $\alpha$ is presented in Fig.5 for two values of $\theta$, both negative and positive.  
Whereas in the former case $\mu$ is almost constant -- and larger than the value predicted by Eq.(\ref{war1t0}) -- 
for the positive $\theta$ we observe a linear growth in a wide range of $\alpha$; 
$\mu(\alpha)$ becomes flat only at large $\alpha$. Finally, Fig.6 shows that the slope for a positive (negative) 
$\theta$ rises with $\beta$ weaker (stronger) than for $\theta=0$ and both dependences are linear. 
The superdiffusion emerges when $\theta$ is negative and $\beta$ large. 

The proposed formalism is applicable to diffusion problems where memory is connected 
with nonhomogeneous medium structure. Therefore, we conclude with some remarks concerning the parameters 
in the above analysis; they may be useful to compare the results with an experiment. 
We interpreted the multiplicative noise $f(x)$ in Eq.(\ref{lam}) according to SI. 
This interpretation is distinguished in stochastic problems because it constitutes a white-noise 
limit of coloured noises for any $\alpha$ \cite{sro12}. However, the other interpretations 
may also be important. For example, the experimental analysis of the colloidal particles 
diffusion near the boundary favours AII ($\lambda_I=1$) \cite{brett}, as we have already mentioned. 
We performed the numerical calculations for AII, similar to those for SI, and found 
the same diffusion properties in respect both to the exponent and the proportionality coefficient. 
This conclusion could be anticipated from Fig.1: tails of the distribution for both interpretations are 
very similar. Moreover, results presented in Fig.4-6 are independent of $\gamma$ and $L$ if $L$ is 
sufficiently large. The other parameters have a straightforward interpretation: 
$\alpha$ defines the jump statistics, $\beta$ is responsible for the trapping time characterising the depth of 
the effective trapping potential and $\theta$ is responsible for the trap distribution.

\section{Summary and conclusions} 

The stochastic motion in a medium with traps was studied in terms of the Langevin equation and 
the anomalous diffusion exponent was determined. 
The memory effects were taken into account by a subordination of a Markovian process to a physical, random 
time and the nonhomogeneity of the trap structure by a dependence of the time lag on the position, modelled by 
a non-negative function $g(x)$. If one decouples effects related to the trap structure and memory, 
the problem resolves itself to a multiplicative process subordinated to the random time. 
The Langevin equation in the operational time describes a Markovian jumping process with a variable 
rate of the Poissonian waiting-time distribution. 
The density distribution can be exactly derived if $g(x)$ has a power-law form. It has 
a stretched-Gaussian shape and this result is similar to the prediction of the quenched 
trap model \cite{bou,ber}. However, in contrast to that model, the variance may rise with time faster than 
linearly; beside the subdiffusion, observed for the homogeneous case, the enhanced diffusion emerges.
If one interprets the position-dependence of the memory in terms of a variable trap density and 
assume it in a power-law form, such a density corresponds to a fractal pattern. Then the fractal dimension 
determines the diffusion properties: the smaller the dimension, the faster the variance grows with time. 

L\'evy flights are characterised by the infinite variance but a finite size of 
the system may modify the random stimulation. In contrast to the usual truncation procedure 
where long jumps are eliminated, our approach imposes a restriction on the noise intensity 
by introducing a multiplicative noise to the Langevin equation. This noise may follow from 
a complicated medium structure near the boundary which requires 
a generalisation of a simple modelling in terms of the additive noise. 
The resulting density distributions have fast falling tails. Therefore, diffusion inside 
the substrate is well-determined and may be quantified in terms of the variance 
the time-dependence of which does not depend of the system size. Moreover, 
the diffusion properties appear robust in respect to a particular interpretation of the multiplicative 
noise in the Langevin equation. On the other hand, diffusion inside a layer near the boundary 
is weaker than in the bulk and we observe two regimes with a different anomalous diffusion 
exponent $\mu$. This exponent reflects both the magnitude of the memory in the system, described by the parameter 
$\beta$, and the position-dependence of the memory, given by the parameter $\theta$: it diminishes 
with $\theta$ and rises with $\beta$. In contrast to the case $\theta=0$, we observe a dependence 
of $\mu$ on $\alpha$ but only for small $\alpha$ and positive $\theta$. This conclusion points at 
a subtle relation between the nonhomogeneous distribution of the random time statistics 
and the noise statistics in respect to the diffusion properties of systems with the L\'evy flights. 

\section*{APPENDIX A}

\setcounter{equation}{0}
\renewcommand{\theequation}{A\arabic{equation}} 

In this Appendix, we demonstrate that the first part of Eq.(\ref{las}) 
traces back to a Markovian jumping process \cite{kam}. 
This stationary process is defined by a jump-size distribution $Q(x)$ in a form of the L\'evy $\alpha$-stable and 
symmetric distribution with a characteristic function, 
\begin{equation}
\label{A.1}
{\widetilde Q}(k)=\exp(-K^\alpha |k|^\alpha)~~~~~~~~~~(\alpha\ne1, K>0).
\end{equation} 
The particle performs instantaneous jumps and then rests for a time given by a waiting-time distribution 
which is Poissonian, 
\begin{equation}
\label{A.2}
w(t)=\nu(x){\mbox e}^{-\nu(x)t}, 
\end{equation}
where $\nu(x)$ denotes a position-dependent rate. The transition probability for infinitesimal time intervals 
$\Delta t$ reads 
\begin{equation}
\label{A.3}
p_{tr}(x,\Delta t|x', 0) = [1-\nu(x')]\Delta t\delta(x-x')+Q(x-x')\nu(x') \Delta t,
\end{equation} 
where the first term corresponds to the case that no jump occurred within $\Delta t$ and the second one that 
exactly one jump occurred. The differentiation over time, 
\begin{equation}
\label{A.4}
\frac{\partial}{\partial t}p(x,t)=\lim_{\Delta t\to 0}\left[\int p_{tr}(x,\Delta t|x',0)
p(x',t)dx' - p(x,t)\right]/\Delta t,
\end{equation} 
produces a master equation: 
  \begin{equation}
  \label{A.5}
  \frac{\partial}{\partial t}p(x,t) = -\nu(x)p(x,t) +
  \int Q(x',x)\nu (x') p(x',t) dx'.
  \end{equation} 
In the diffusion limit $k\to0$, Eq.(\ref{A.1}) can be approximated by 
${\widetilde Q}(k)\approx 1-K^\alpha |k|^\alpha$ 
and inserting this expression into the Fourier transformed Eq.(\ref{A.5}) yields \cite{sro06}
\begin{equation}
\label{A.6}
\frac{\partial{\widetilde p}(k,t)}{\partial t}=-K^\alpha |k|^\alpha{\cal F}[\nu(x)p(x,t)]. 
\end{equation} 
Inversion of the above equation yields a fractional Fokker-Planck equation in the form 
\begin{equation}
\label{A.7}
\frac{\partial p(x,t)}{\partial t}=K^\alpha\frac{\partial^\alpha[\nu(x)p(x,t)]}{\partial|x|^\alpha},
\end{equation} 
where the Weyl-Riesz operator is defined by the inverse Fourier transform: 
$\frac{\partial^\alpha}{\partial|x|^\alpha}={\cal F}^{-1}(-|k|^\alpha)$. On the other hand, Eq.(\ref{A.7}) 
follows from the first of Eq.(\ref{las}) in the It\^o interpretation \cite{sche}. 

\section*{APPENDIX B}

\setcounter{equation}{0}
\renewcommand{\theequation}{A\arabic{equation}} 

In Appendix B, we derive Eq.(\ref{sol2}). The Bessel function ${\mbox K}_\nu(z)$ can be expressed 
in terms of the Fox H-function \cite{kil} which formula, in our case, reads
\begin{eqnarray} 
\label{B.1}
{\mbox K}_\nu(z)=\frac{1}{2}H_{0,2}^{2,0}\left[\frac{z^2}{4}\left|\begin{array}{l}
~~~~~~-\!\!\!-\!\!\!-\!\!\!-\!\!\!-\!\!\!-\!\!\!-\\
\\
(-\nu/2,1),(\nu/2,1)
\end{array}\right.\right]. 
\end{eqnarray} 
After applying standard properties of the H-function and straightforward calculations 
we obtain the density in the form 
\begin{eqnarray} 
\label{B.2}
\bar p(x,u)=-\frac{1}{\beta}\frac{(1+\theta/2)^{\nu+2c/\beta}}{\Gamma(-\nu)}|x|^{(2+\theta)/\beta-1}
H_{0,2}^{2,0}\left[\xi\left|\begin{array}{l}
~~~~~~~~~~~~~~~~-\!\!\!-\!\!\!-\!\!\!-\!\!\!-\!\!\!-\!\!\!-\\
\\
(c/\beta-\nu/2,1/\beta),(c/\beta+\nu/2,1/\beta)
\end{array}\right.\right], 
\end{eqnarray} 
where $\xi=(2+\theta)^{-2/\beta}|x|^{(2+\theta)/\beta}u\equiv\kappa u$. Inversion of the transform 
enhances the order of the H-function \cite{glo}; in the case of the function in Eq.(\ref{B.2}), 
the inversion formula yields 
 \begin{eqnarray} 
\label{B.3}
\frac{1}{t}
H_{1,2}^{2,0}\left[\frac{\kappa}{t}\left|\begin{array}{l}
~~~~~~~~~~~~~~~~~~~(0,1)\\
\\
(c/\beta-\nu/2,1/\beta),(c/\beta+\nu/2,1/\beta)
\end{array}\right.\right]. 
\end{eqnarray} 
Finally, we obtain Eq.(\ref{sol2}).


\begin{thebibliography}{99}
\bibitem{uwa}
The asymptotics $|y|^{-1-\alpha}$ is valid, in particular, for any positive $\theta$ but the approximation 
of the solution of Eq.(\ref{fp0}) near the origin by the L\'evy stable distribution imposes 
an upper limit on $\theta$ \cite{sro14a}. The solution for large $\theta$ requires a modification 
of the stable form \cite{sro06}.
\end{thebibliography}
\end{document}